\documentclass{emulateapj}

\usepackage{amsmath}
\usepackage{amssymb}

\newcommand{\ra}[4]{${#1}^{\rm h}{#2}^{\rm m}{#3}\fs{#4}$}
\newcommand{\dec}[4]{${#1}\arcdeg{#2}\arcmin{#3}\farcs{#4}$}

\slugcomment{Submitted to ApJ.}

\shorttitle{A Neptune-mass Free-floating Planet Candidate Discovered by Microlensing Surveys}
\shortauthors{Mr\'oz et al.}

\begin{document}

\title{A Neptune-mass Free-floating Planet Candidate Discovered by Microlensing Surveys}

\author{Przemek Mr\'oz$^{1}$\altaffilmark{$\dagger$}, Y.-H. Ryu$^{2}$, J. Skowron$^{1}$, A. Udalski$^{1}$, A. Gould$^{2,3,4}$, \\and\\ M.~K. Szyma\'nski$^{1}$, I. Soszy\'nski$^{1}$, R. Poleski$^{3}$, P. Pietrukowicz$^{1}$, \\S. Koz\l{}owski$^{1}$, M. Pawlak$^{1}$, K. Ulaczyk$^{5}$ \\(The OGLE Collaboration), \\and\\ M.~D.~Albrow$^{6}$, S.-J.~Chung$^{2,7}$, Y.~K.~Jung$^{8}$, C.~Han$^{9}$, K.-H.~Hwang$^{2}$, I.-G.~Shin$^{8}$, J.~C.~Yee$^{8}$, W.~Zhu$^{10}$, S.-M.~Cha$^{2,11}$, D.-J.~Kim$^{2}$, H.-W.~Kim$^{2}$, S.-L.~Kim$^{2,7}$, C.-U.~Lee$^{2,7}$, D.-J.~Lee$^{2}$, Y.~Lee$^{2,11}$, B.-G.~Park$^{2,7}$, and R.~W.~Pogge$^{3}$ \\(The KMTNet Collaboration)}


\affil{$^{1}$Warsaw University Observatory, Al. Ujazdowskie 4, 
00-478 Warszawa, Poland}
\affil{$^{2}$Korea Astronomy and Space Science Institute, Daejon
34055, Republic of Korea}
\affil{$^{3}$Department of Astronomy, Ohio State University, 140 West
18th Avenue, Columbus, OH 43210, USA}
\affil{$^{4}$Max-Planck-Institute for Astronomy, K\"{o}nigstuhl 17,
D-69117 Heidelberg, Germany}
\affil{$^{5}$Department of Physics, University of Warwick, Coventry
CV4 7AL, UK}
\affil{$^{6}$University of Canterbury, Department of Physics and
Astronomy, Private Bag 4800, Christchurch 8020, New Zealand}
\affil{$^{7}$Korea University of Science and Technology, 217 Gajeong-ro, 
Yuseong-gu, Daejeon, 34113, Republic of Korea}
\affil{$^{8}$Harvard-Smithsonian Center for Astrophysics, 60 Garden
Street, Cambridge, MA 02138, USA}
\affil{$^{9}$Department of Physics, Chungbuk National University,
Cheongju 28644, Republic of Korea}
\affil{$^{10}$Canadian Institute for Theoretical Astrophysics,
University of Toronto, Toronto, ON M5S 3H8, Canada}
\affil{$^{11}$School of Space Research, Kyung Hee University,
Yongin, Kyeonggi 17104, Republic of Korea}

\altaffiltext{$\dagger$}{Corresponding author: pmroz@astrouw.edu.pl}

\begin{abstract}

Current microlensing surveys are sensitive to free-floating planets down to Earth-mass objects. All published microlensing events attributed to unbound planets were identified based on their short timescale (below two days), but lacked an angular Einstein radius measurement (and hence lacked a significant constraint on the lens mass). Here, we present the discovery of a Neptune-mass free-floating planet candidate in the ultrashort ($t_{\rm E}=0.320\pm0.003$ days) microlensing event OGLE-2016-BLG-1540. The event exhibited strong finite-source effects, which allowed us to measure its angular Einstein radius of $\theta_{\rm E}=9.2\pm0.5\,\mu$as. There remains, however, a degeneracy between the lens mass and distance. The combination of the source proper motion and source-lens relative proper motion measurements favors a Neptune-mass lens located in the Galactic disk. However, we cannot rule out that the lens is a Saturn-mass object belonging to the bulge population. We exclude stellar companions up to $\sim 15\,$au.

\end{abstract}


\section{Introduction}

Gravitational microlensing enables detecting dark objects in a broad mass range: from black holes and neutron stars to Earth-sized planets \citep{paczynski}. In particular, current microlensing surveys are sensitive to free-floating planets, which are gravitationally unbound to any star. A characteristic timescale of a microlensing event (known as the Einstein time $t_{\rm E}$) depends on the relative lens-source proper motion $\mu_{\rm rel}$ and the angular Einstein radius $\theta_{\rm E}$:

\begin{align}
t_{\rm E} &= \frac{\theta_{\rm E}}{\mu_{\rm rel}} = \frac{\sqrt{\kappa M \pi_{\rm rel}}}{\mu_{\rm rel}} 
\label{eq:tE}
\end{align}
where $M$ is the lens mass, $\pi_{\rm rel}=1\ \textrm{au}\,(D^{-1}_{\rm L} - D^{-1}_{\rm S})$ is the lens-source relative parallax, and $\kappa = 4G/(c^2\,\textrm{au}) = 8.14\ \textrm{mas}/M_{\sun}$. Here, $D_{\rm L}$ and $D_{\rm S}$ are distances to the lens and source, respectively.

The Einstein timescale is the only physical parameter that can be measured for the majority of microlensing events. As the timescale is proportional to the square root of mass, it is expected that events caused by free-floating planets are very short ($t_{\rm E} \lesssim 2\,\mathrm{days}$). However, the mass measurement requires the knowledge of two additional physical parameters: the angular Einstein radius $\theta_{\rm E}$ and the microlens parallax $\pi_{\rm E} = \pi_{\rm rel}/\theta_{\rm E}$.

Although the angular Einstein radius is routinely measured in binary microlensing events via the finite-source effect \citep{udalski1994,mao1994,nemi1994}, such a measurement is much harder for single lensing events because it requires that the source passes almost exactly over the lens to produce a detectable finite-source signal \citep{alcock1997,yoo2004}. It is expected that finite-source effects should be strong for Earth-mass lenses, because the angular size of the source is comparable to the Einstein ring radius \citep{bennett1996,mao2016}. To date, however, no such measurements have been reported.

The microlens parallax measurements are even harder for free-floating planets. The subtle deviations from the standard microlensing light curve due to parallax can be detected in long-timescale events, as the Earth-based observer moves along the orbit \citep{gould1992}. Parallax can be also measured using simultaneous ground- and space-based observations \citep{refsdal1966}, for example, with the \textit{Spitzer} satellite \citep{dong2007,udalski-spitzer}. However, \textit{Spitzer} operations require the targets to be uploaded to the spacecraft at least three days in advance, making observations of short events nearly impossible. The problem can be overcome with continuous, survey-mode observations (e.g., \citealt{henderson2016,gould2016}). Such an experiment was conducted during the \textit{K2} Campaign 9 \citep{henderson,penny2017,zhu2017}, but owing to the difficulties in extracting the photometry from crowded regions of the Galactic bulge, no observations of short-timescale microlensing events from \textit{K2}C9 were reported so far.

Information about the mass function of lenses, including free-floating planets, can be inferred from a statistical analysis of the distribution of timescales of a large sample of microlensing events. The first such measurement was attempted by \citet{sumi2011}, who analyzed a sample of 474 microlensing events detected by the Microlensing Observations in Astrophysics (MOA) group. They found an excess of nine short\footnote{They reported ten events with timescales shorter than 2 days, but one measurement, for MOA-ip-1, is incorrect \citep{mroz}. } events, relative to what was expected from brown dwarfs and stars, and they attributed this excess to a large population of Jupiter-mass planets, which should be nearly twice as common as main-sequence stars. 

\citet{clanton} modeled the microlensing signal expected from exoplanets on wide orbits using constraints from microlensing, radial velocity, and direct imaging surveys and concluded that at most $\sim40\%$ of short-timescale events detected by \citet{sumi2011} can be interpreted as due to wide-orbit planets. However, the statistical significance of Sumi et al.'s results is largely based on three shortest-timescale events ($t_{\rm E}<1$ days). As we mentioned above, one measurement is incorrect and the model by \citet{clanton} shows that the remaining two are statistically consistent with being wide-orbit planets. That model still cannot account for a small overabundance of events with timescales between 1--2 days (see Figures 4 and 5 from \citealt{clanton}), but the statistical significance of the remaining excess relative to the short-timescale events expected from stars and brown dwarfs is small.

A large population of Jupiter-mass free-floating planets suggested by \citet{sumi2011} was difficult to reconcile with censuses of substellar objects in young clusters and star-forming regions and with predictions of planet-formation theories. For example, \citet{pena2012} and \citet{scholz2012} analyzed substellar mass functions of the young clusters $\sigma$ Orionis and NGC~1333, finding that free-floating planetary-mass objects are at least an order of magnitude less common than main-sequence stars. These observations are incomplete for masses below $\sim 6\,M_{\rm Jup}$, so direct comparisons with microlensing surveys are difficult. Several mechanisms of free-floating planet production have been proposed (e.g., \citealt{veras2012}), but none of them is capable of explaining the large number of Jupiter-mass free-floaters suggested by \citeauthor{sumi2011} On the other hand, Earth- and super-Earth-mass planets can be scattered and ejected much more efficiently \citep{pfyffer2015,mao2016,barclay2017}.

The recent analysis of data from the Optical Gravitational Lensing Experiment (OGLE) \citep{udalski2015} provides much stronger constraints on the abundance of free-floating Jupiters. \citet{mroz} analyzed a larger sample of over 2,600 microlensing events discovered during the years 2010--2015. They found that Jupiter-mass lenses are at most an order of magnitude less common than suggested by \citeauthor{sumi2011} (with a 95\% upper limit of 0.25 Jupiter-mass planets per main-sequence star). They detected, however, a handful of ultrashort-timescale microlensing events (with timescales of less than 0.5 day), strongly suggesting the existence of Earth-mass and super-Earth-mass free-floating planets. Their light curves are not well covered with observations from a single telescope, rendering the detection of the finite-source effect difficult. 

We conducted a pilot program of searching for ultrashort microlensing events in the 2016 observing season data. We supplemented OGLE observations with data from the KMTNet survey, a network of longitude-separated telescopes, which provided us with a better coverage of short-timescale microlensing events.  

Here, we present the discovery of an ultrashort-timescale event OGLE-2016-BLG-1540 and report the first measurement of the Einstein ring radius of a free-floating planet candidate. 

\section{Observations}
\label{sec:data}

Microlensing event OGLE-2016-BLG-1540 was discovered by the OGLE Early Warning System \citep{udalski2003} on 2016 August 6, at equatorial coordinates of R.A. = \ra{18}{00}{47}{00} and Decl. = \dec{-28}{21}{35}{2} (J2000.0), i.e., Galactic coordinates $l=2.186^{\circ}$, $b=-2.574^{\circ}$. The survey uses a 1.3~m Warsaw Telescope at Las Campanas Observatory in Chile (the Observatory is operated by the Carnegie Institution for Science), equipped with a 1.4 deg$^2$ mosaic CCD camera. The event was located in field BLG512, which was observed with a cadence of 20 minutes. 

The Korea Microlensing Telescope Network (KMTNet) consists of three 1.6~m telescopes equipped with 4.0 deg$^2$ cameras. The telescopes are located in CTIO (Chile), SAAO (South Africa), and SSO (Australia), see \citet{kim2016} for details. The event was located in two overlapping fields BLG03 and BLG43, monitored with a cadence of 14 minutes. We omitted KMT~SSO observations, because they did not cover the peak and did not contribute to constraining the model. We also excluded KMT~SAAO data taken before August 3 or after August 16, because the baseline light curve showed systematic variability connected with passages of the Moon near the bulge fields.

All observations used in the modeling were taken in the $I$ band. Photometry was extracted using custom implementations of the difference image analysis \citep{alard}: \citet{wozniak} (OGLE) and \citet{albrow2009} (KMTNet). The photometric uncertainties were corrected using the standard procedures \citep{skowron2016}. We additionally reduced KMT~CTIO $V$- and $I$-band images using DoPhot \citep{schechter1993}, which allowed us to determine the source color. 

\begin{figure*}
\centering
\includegraphics[width=0.7\textwidth]{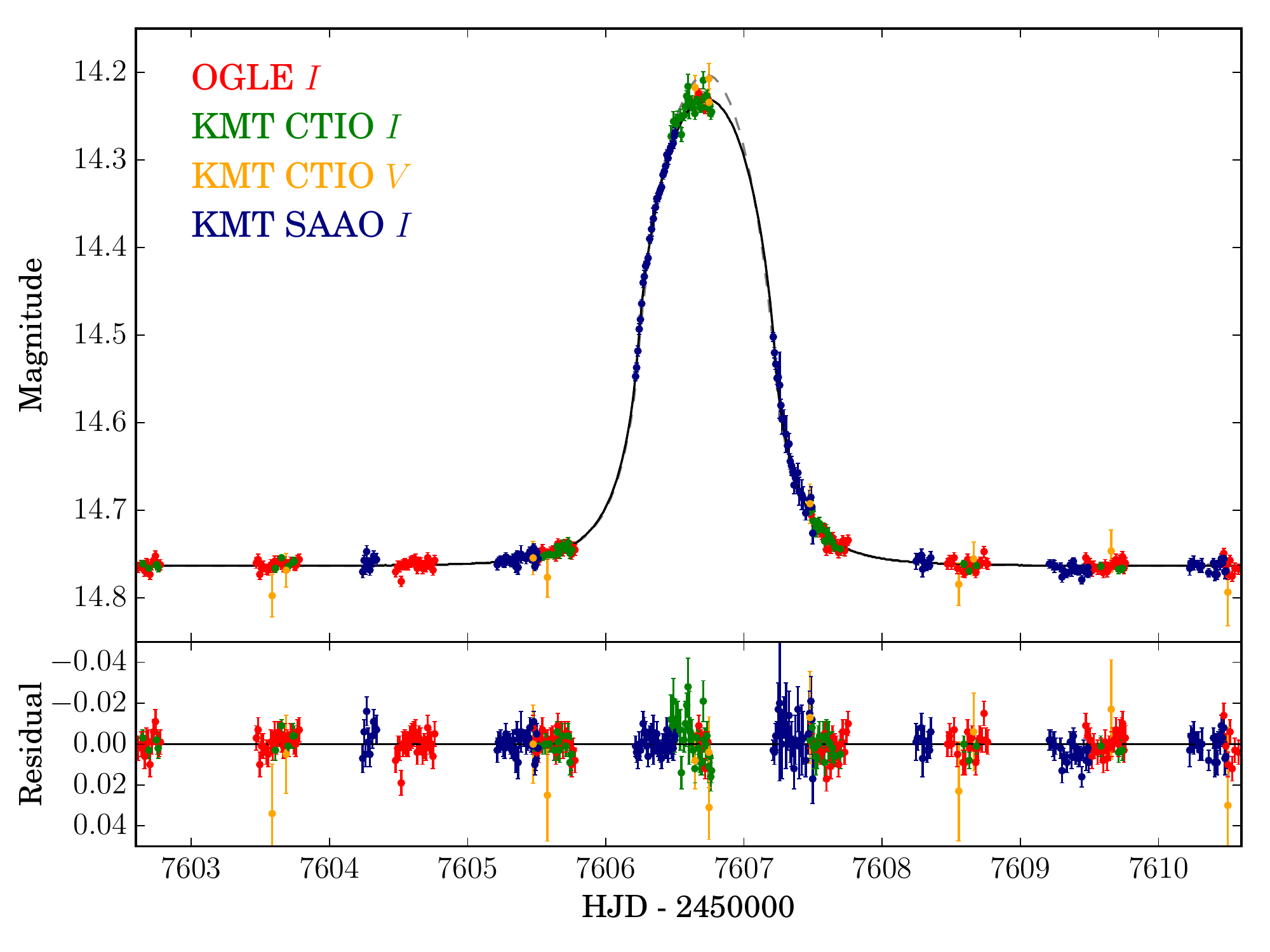}
\caption{Microlensing event OGLE-2016-BLG-1540 exhibits prominent finite-source effect, because the source is larger than the angular Einstein ring. The light curve can be accurately described using the finite-source point-lens model (black solid line in the $I$-band, gray dashed line in the $V$-band). $I$- and $V$-band models differ because of different limb-darkening profiles of the source star in two filters. $V$-band data were not used in the modeling. All measurements were transformed to the OGLE magnitude scale.}
\label{fig:lc}
\end{figure*}

\section{Light Curve Modeling}
\label{sec:model}

The light curve of the event (Figure \ref{fig:lc}) can be accurately described using the finite-source point-lens model. The model has four parameters: the time and projected separation of closest approach of the source to the lens $t_0$ and $u_0$, the Einstein timescale $t_{\rm E}$, and the normalized angular radius of the source $\rho = \theta_* / \theta_{\rm E}$ ($\theta_*$ is the angular radius of the source). 

Two parameters, $F_{\rm s}$ and $F_{\rm b}$, describe the source and (unmagnified) blend fluxes, respectively. When we allowed $F_{\rm b}$ to vary, we found that in the best-fit solution the blend flux is negative (with the absolute value corresponding to a 16--17-mag star). Although such solutions are mathematically possible, this negative blending is too big to be due to normal fluctuations in the background. The best-fit solution is only $\Delta\chi^2=5$ better than the solution with fixed $F_{\rm b}=0$, which can easily be due to statistical noise, or possibly low-level systematics in the data. Given the absence of evidence for blending and the low prior probability for ambient superposed bright source, our best estimate for the blended light is zero, i.e., $F_{\rm b}=0$. The only way that the source flux enters the characterization of the lens is via $\theta_*$ (see Section \ref{sec:source}). To account for this, while we fix $F_{\rm b}=0$ in the fits, we also add in quadrature 0.05 mag to the uncertainty in centroiding the clump, when we compute our errors of these quantities.

Two additional (wavelength-dependent) parameters $\Gamma$ and $\Lambda$ may be used to describe the limb-darkening profile: $S(\varphi)/\bar{S} = 1 - \Gamma (1-\frac{3}{2}\cos\varphi) - \Lambda (1-\frac{5}{4}\sqrt{\cos\varphi})$, where $\varphi$ is the angle between the normal to the stellar surface and the line of sight \citep{yoo2004}. The two-parameter limb darkening law provides a more accurate description of a brightness profile than a simple linear law (e.g., \citealt{albrow1999,fields2003,abe2003}).
We used a fixed $\Gamma_I = 0.36$ and $\Lambda_I=0.34$ which correspond to the physical parameters of the source star (c.f., Section \ref{sec:source}). When we allowed $\Gamma$ and $\Lambda$ to vary, we found $\Gamma=0.25 \pm 0.20$ and $\Lambda=0.36 \pm 0.40$, consistent at $1.5-2\sigma$ level with the adopted values. 

The finite-source magnifications were calculated by the direct integration of formulae derived by \citet{lee2009}, which remain valid in the low-magnification regime. The uncertainties were estimated using the Markov Chain Monte Carlo method. The best-fitting parameters and their $1\sigma$ error bars are shown in Table \ref{tab:pars}.

We also considered models with terrestrial parallax \citep{gould2009,freeman2015}. The microlens parallax in the best-fitting solution was $\pi_{\rm E}=3200 \pm 700$, but the $\chi^2$ improvement was modest ($\Delta\chi^2=18$). The parallax signal came mostly from one observatory (KMT~CTIO) from one night and the OGLE data from that night did not provide strong evidence for parallax. Thus, the terrestrial parallax signal may be mimicked by some low-level systematics in the data and cannot be trusted.

\begin{table}
\caption{Best-fitting model parameters}
\centering
\begin{tabular}{lcc}
\hline
Parameter & Value & Uncertainty \\
\hline
$t_0$ (HJD$'$) & 7606.726 & 0.002 \\
$t_{\rm E}$ (days) & 0.320 & 0.003 \\
$u_0$ & 0.53 & 0.04 \\
$\rho$ & 1.65 & 0.01 \\
\hline
$I_{\rm s}$ & 14.76 & 0.05 \\
$f_{\rm s}$ & 1.00 & (fixed) \\
\hline
$\chi^2/\mathrm{d.o.f.}$ & \multicolumn{2}{c}{2160.1/2153}\\
\hline
\end{tabular}

Note: HJD$'$=HJD-2450000. $f_{\rm s}=F_{\rm s}/(F_{\rm s}+F_{\rm b})$ is the blending parameter.
\label{tab:pars}
\end{table}

\section{Physical Parameters}
\label{sec:physical}

\subsection{Source star}
\label{sec:source}

\begin{figure}
\centering
\includegraphics[width=0.5\textwidth]{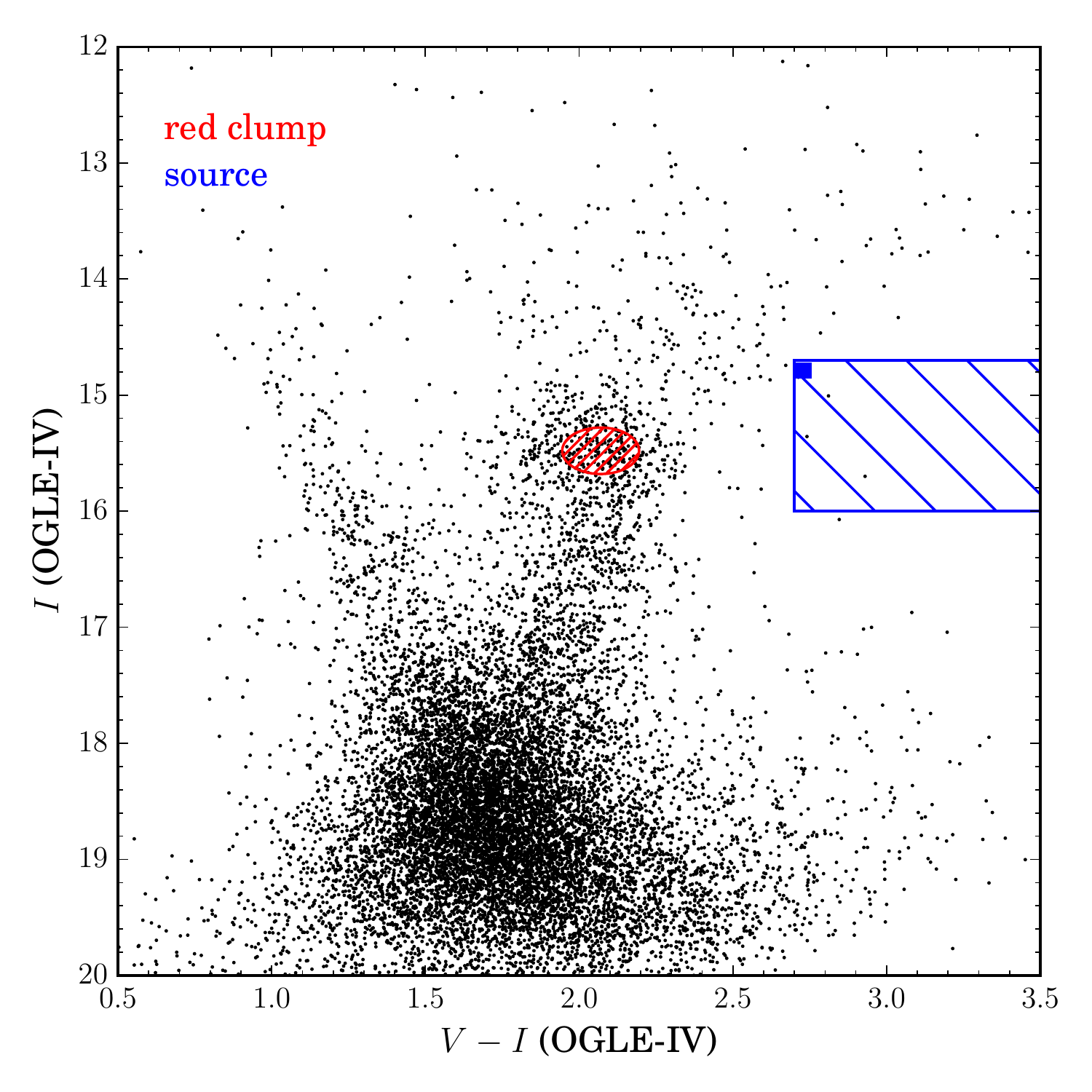}
\caption{The OGLE-IV color--magnitude diagram for stars in a $4' \times 4'$ region around OGLE-2016-BLG-1540. The source (blue square) is located in a relatively lowly populated region of the diagram. Blue and red areas mark stars used for the proper motion measurements (see Section \ref{sec:pm}).}
\label{fig:cmd}
\end{figure}

The event was observed in the $V$-band by the KMT~CTIO on the peak night (Figure \ref{fig:lc}), which allowed us to measure the color of the source and hence the Einstein ring radius $\theta_{\rm E}$ \citep{yoo2004}. Because the finite-source effect is prominent and the event may no longer be achromatic, we have not used the model-independent regression to estimate the source color. Instead, we calculated the color for each link of the MCMC chain. The procedure of model fitting, calculating the source color and the limb-darkening coefficients was iterated, until the color measurement converged. We compared the location of the source and red clump centroid in the instrumental color--magnitude diagram (CMD) in a $2' \times 2'$ region around the event for KMT~CTIO. We found that the source is $\Delta(V-I)=0.61 \pm 0.02$ redder and $\Delta I = -0.85 \pm 0.09$ brighter than the red clump. Assuming the intrinsic color of $(V-I)_{\rm{RC},0}=1.06$ of red clump stars \citep{bensby2011} and their mean de-reddened brightness in this direction of $I_{\rm RC,0}=14.36$ \citep{nataf2013}, we calculated the intrinsic brightness $I_{\rm S,0}=13.51 \pm 0.09$ and color $(V-I)_{\rm S,0} = 1.67 \pm 0.02$ of the source. The OGLE-IV CMD, for a larger region of $4'\times 4'$, is shown in Figure \ref{fig:cmd}.

We then found $(V-K)_{\rm S,0} = 3.67 \pm 0.03$ from color--color relations from \citet{bessel1988} and estimated the angular radius of the source star $\theta_* = 15.1 \pm 0.8$~$\mu$as from color-surface brightness relation for giants \citep{kervella2004}. The latter estimate allowed us to measure the angular Einstein radius 
$$
\theta_{\rm E} = \theta_{*} / \rho = 9.2 \pm 0.5\,\mu\mathrm{as} 
$$
and the relative lens-source proper motion 
$$
\mu_{\rm rel} = \theta_{\rm E} / t_{\rm E} = 10.5 \pm 0.6\,\mathrm{mas\,yr}^{-1}.
$$

We can also estimate the effective temperature of the source of $T_{\rm eff} = 3900 \pm 200$~K using the color--temperature relations of \citet{houdashelt2000} and \citet{ramirez2005}. The corresponding limb-darkening coefficients \citep{claret2011} are:

\begin{tabular}{r@{=}lr@{=}l}
$\Gamma_I$ & $0.36$ & $\Lambda_I$ & $0.34$ \\
$\Gamma_V$ & $0.94$ & $\Lambda_V$ & $-0.21$
\end{tabular}

We used ATLAS models and assumed a solar metallicity, microturbulent velocity of 2 km/s and surface gravity of $\log\,g=2.0$. Limb-darkening coefficients $(c,d)$ from \citet{claret2011} were transformed to $(\Gamma,\Lambda)$ using formulae derived by \citet{fields2003}.

\subsection{Source proper motion}
\label{sec:pm}

As the source star is relatively bright and the contribution of the lens to the total light is negligible, it is possible to measure the absolute proper motion of the source. We chose a subset of 363 best-seeing ($0.7-1''$) and low-background images (out of 11,276 total epochs of this field) spanning 2010--2017 taken with the 24th CCD detector of the OGLE-IV camera. We used the CMD to identify 3818 candidate red clump stars, which served as anchors for the coordinates transformations between the CCD frames. This allowed us to measure proper motions with respect to the mean motion of the Galactic bulge.

\begin{figure}
\centering
\includegraphics[width=0.5\textwidth]{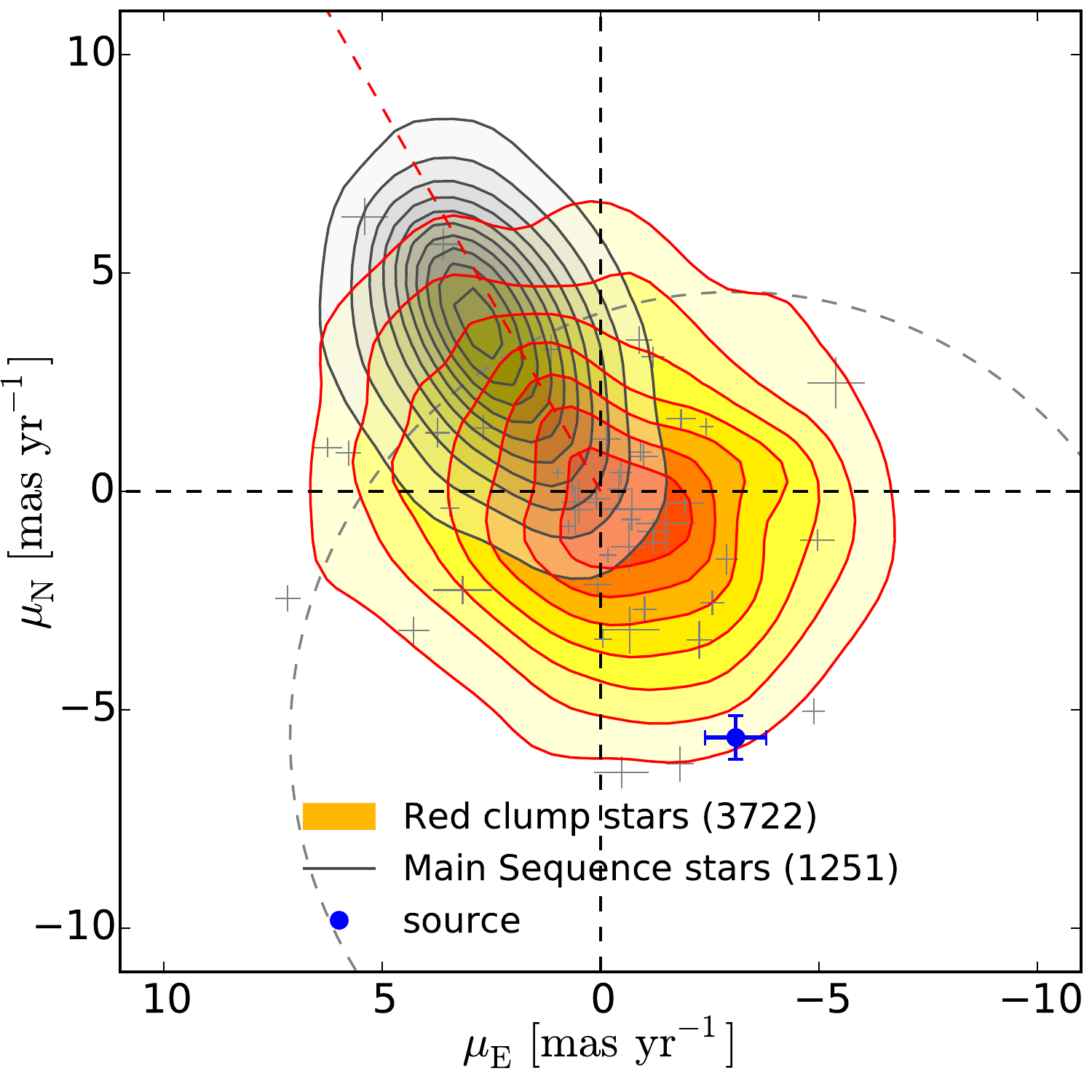}
\caption{Proper motions of stars in the OGLE-2016-BLG-1540 field ($9'\times 18'$). Orange contours mark proper motions of red clump stars (bulge population), black contours mark main-sequence stars (which represent the Galactic disk population). The dashed red line shows the direction of increasing Galactic longitude and the proper motion of the source star is marked with a blue dot. The gray dashed circle corresponds to the relative proper motion of 10.5 mas\,yr$^{-1}$ with respect to the source. As the lens should be located on this circle, it likely belongs to the Galactic disk population. Gray crosses mark stars located near the source in the CMD; they follow the bulge distribution.}
\label{fig:pm}
\end{figure}

We used the DoPhot PSF photometry package \citep{schechter1993} to measure positions of all stars in all 363 images. Then, we calculated the third-order polynomial coordinate transformations between each frame and the first frame by minimizing the scatter for the anchor red clump stars. The proper motions were fitted with the least-squares method (with outlier rejection). The formal uncertainties of the fit were typically $0.2-0.3$ mas\,yr$^{-1}$. However, the comparison with proper motion measurements based on the OGLE-III data (2001--2009) showed discrepancies larger than the pure statistical error. We decided to employ 0.5 and 0.7 mas\,yr$^{-1}$ for N and E directions as our measure of uncertainty; hence, the proper motion of the source is $\boldsymbol{\mu}_{\rm S} = (\mu_{\rm N}, \mu_{\rm E}) = (-5.6 \pm 0.5,  -3.0 \pm 0.7)\,\rm{mas\,y}r^{-1}$ with respect to the Galactic bulge (see Figure \ref{fig:pm}). 

As both the position of the source star on the CMD is uncommon (the star is located below the red giant branch and redwards of red clump giants) and its proper motion is counter to the Galactic rotation, we consider whether this evidence indicates that the source belongs to the far disk population. First, we investigated the CMD position of the source. We identified about 45 stars with similar CMD positions\footnote{The analyzed region is marked with a blue rectangle in Figure \ref{fig:cmd}.} and measured their kinematics (their proper motions are marked with gray crosses in Figure \ref{fig:pm}), finding that these are consistent with all other red giant (i.e., bulge) stars in the field\footnote{We compared both distributions of proper motions using the two-sample Anderson--Darling test and found $p$-values of 0.21 (for $\mu_l$ component) and 0.65 ($\mu_b$). Similarly, the two-sample Kolmogorov--Smirnov test yields $p$-values of 0.29 and 0.34, respectively. Therefore, there is no evidence that these distributions are different.}. Thus, the unusual position of the source star on the CMD cannot be taken as evidence for belonging to some other population. 

\subsection{Constraints on the host star}

Additional features in the event light curve would have been detected if the trajectory of the source were fortunate enough to pass near a putative host star. Because the light curve does not exhibit any signatures of the host star, we can only provide a lower limit on the planet-host separation, using a variation of the method proposed by \citet{gaudi2000}. We simulated artificial OGLE light curves (spanning from 2010 March 4, through 2017 October 10) for a given binary model, defined by three additional parameters as compared to the single-lens case: mass ratio $q$, star-planet separation $s$ (expressed in Einstein radii), and angle $\alpha$ between the source trajectory and binary axis. (The remaining parameters were calculated based on the best-fitting single-lens model from Table \ref{tab:pars}). We fitted binary and single lensing models to the artificial data and calculated the difference $\Delta\chi^2 = \chi^2_{\rm single}-\chi^2_{\rm binary}$. We used $q=2\times 10^{-4}$, which corresponds to the Einstein radius of the host of $\theta_{\rm E,host}=\theta_{\rm E}/\sqrt{q}=0.65\,$mas. If the lens is located in the Galactic disk ($\pi_{\rm rel}=0.1$\,mas), the corresponding host mass is $M_{\rm host}=0.5\,M_{\odot}$. For each value of $s$, we simulated 180 light curves with uniformly distributed $\alpha\in[0,2\pi]$, and calculated the probability of detecting the host star as the fraction of light curves which fulfill $\Delta\chi^2 > \Delta\chi^2_{\rm thresh} = 225$. This probability drops below 90\% when $s>5.1$. The lower limit on the host separation is slightly weaker for larger mass ratios (because the host event is shorter). For $q=10^{-3}$, corresponding to the $0.5\,M_{\odot}$ host in the bulge ($\pi_{\rm rel}=0.01$\,mas), we found $s>4.8$.

We note that the presence of a putative host may also be revealed by perturbations to the point-lens light curve due to the planetary caustic caused by the central star \citep{han2003a,han2005}. The angular size of the planetary caustic (relative to the Einstein radius of the planet) is $4/s^2\leq 0.16$ for $s\geq 5$. Because the source star is $\sim 10$ times larger, the signatures of the caustic are washed out by the finite-source effect.

\section{Discussion}
\label{sec:discuss}

Current microlensing surveys are capable of detecting free-floating planets down to Earth-mass objects. To this day, however, all reported free-floating planet candidates were based on the very short timescale of an event ($t_{\rm E} \lesssim 2$ days) and lacked direct measurements of the angular Einstein ring size \citep{sumi2011,mroz}. OGLE-2016-BLG-1540 is the first case for which we procured such a measurement, owing to the fortuitous fact that the source was a giant. If the source were a dwarf (with at least ten times smaller angular radius), as in the case of ultrashort candidate events detected by \citet{mroz}, the finite-source effect would be significantly weaker. We simulated the OGLE light curve and found that the finite-source model would be preferred only by $\Delta\chi^2 = 1.6$ over the point-lens model.

The short timescale of the event can be explained in part by the unusual kinematics of the system (see Figure \ref{fig:pm}). 
The source is moving at $\mu_{\rm S}=6.4\,\mathrm{mas\,yr}^{-1}$ in the direction opposite to the Galactic rotation and the relative lens-source proper motion is large ($\mu_{\rm rel} = 10.5 \pm 0.6\,\mathrm{mas\,yr}^{-1}$). One possible explanation is that the source is located behind the Galactic center in the far disk, in which case we expect the proper motion direction to be opposite compared to closer stars. To test the ``far disk'' hypothesis, we have studied proper motions of stars located near the source in the CMD (Section~\ref{sec:pm}). These stars follow exactly the same distribution of proper motions as the bulge stars. It appears that, although the source proper motion has an unusual direction, the source belongs to the bulge population.

The large lens-source proper motion indicates that the lens is moving in the opposite direction than the source (along the Galactic rotation) at $\mu_{\rm L} \gtrsim 5$ mas\,yr$^{-1}$ relative to red clump stars. The gray dashed circle in Figure \ref{fig:pm} marks the relative proper motion of $\mu_{\rm rel}=10.5$ mas\,yr$^{-1}$ with respect to the source. As the lens should be located on this circle, it likely belongs to the Galactic disk population. Only 15\% of bulge stars (58\% of the disk stars) are located outside the dashed circle in Figure \ref{fig:pm}, i.e., their proper motions with respect to the source star are higher than $\mu_{\rm rel}$. 

Because the distance to the lens, and so the relative parallax $\pi_{\rm rel}$, is unknown, we cannot uniquely measure the lens mass (eq. (\ref{eq:tE})):
$$
M = \frac{\theta_{\rm E}^2}{\kappa\pi_{\rm rel}} = 35\,M_{\oplus}\frac{0.1\,\mathrm{mas}}{\pi_{\rm rel}}
$$
If the lens is located in the disk ($\pi_{\rm rel}\approx 0.1$~mas), it should be a Neptune-mass planet. However, we cannot rule out that the lens belongs to the bulge population ($\pi_{\rm rel}\approx 0.02$~mas), in which case it should be a Saturn-mass object. The geometry required for the lens to be a brown dwarf ($\pi_{\rm rel}\lesssim 10^{-3}$~mas; i.e., $D_{\rm S}-D_{\rm L} \la 60\,$pc) would require significant fine-tuning. We note that the event occurred inside the \textit{K2}C9 superstamp \citep{henderson}, but unfortunately the parallax measurement was impossible, because the \textit{K2}C9 campaign finished a few weeks before the event.

Microlensing alone cannot distinguish between wide-orbit and unbound planets and, in principle, the lens may be located at a wide orbit, like Uranus or Neptune. 
Our lower limit for the planet-host separation is $5.1$ Einstein radii, which corresponds to the projected physical separation of $15\,\mathrm{au}$ at $\pi_{\rm rel}=0.1$~mas. Owing to the relatively large lens-source proper motion, any stellar companions to the lens can be detected in the future, when the lens and source separate. However, the brightness of the source will make detection of the putative host light difficult.

The characterization of this event would have been impossible without nearly continuous observations from the OGLE and KMTNet surveys. This event also shows the importance of securing the color information for short-timescale events and anomalies (see discussion in \citealt{hwang2017}). Although we could not have measured the precise mass of the lens, such measurements will be possible in the future with the \textit{Euclid} \citep{penny2013} and \textit{WFIRST} \citep{spergel2015}  satellites, but will require simultaneous ground- and space-based observations \citep{gould2013,yee2013,zhu2016}. Current ground-based experiments are already sensitive to ultrashort microlensing events, but a bigger sample is needed to fully understand their origin.

\section*{Acknowledgements}

We thank the anonymous referee for constructive comments. The OGLE project has received funding from the National Science Center, Poland, grant MAESTRO 2014/14/A/ST9/00121 to A.U. Work by W.Z., Y.K.J., and A.G. were supported by AST-1516842 from the US NSF. W.Z., I.G.S., and A.G. were supported by JPL grant 1500811. This research has made use of the KMTNet system operated by the Korea Astronomy and Space Science Institute (KASI) and the data were obtained at three host sites of CTIO in Chile, SAAO in South Africa, and SSO in Australia. Work by K.H.H. was support by KASI grant 2017-1-830-03. Work by C.H. was supported by the grant (2017R1A4A101517) of
National Research Foundation of Korea.

\bibliographystyle{aasjournal}
\bibliography{pap}

\end{document}